# Nanosilica mops up host lipids and fights baculovirus: a *B. mori* model

**Ayesha Rahman[1,2], Dipankar Seth[1], Nitai Debnath[1], C. Ulrichs[2], I. Mewis[2], R. L. Brahmachary[3] and A. Goswami[1]**

[1]Biological Sciences Division, Indian Statistical Institute, 203 B.T. Road, Kolkata- 700 108, West Bengal, India.

[2]Humboldt-Universität zu Berlin, Institut für Gartenbauwissenschaften, Fachgebiet Urbaner Gartenbau, Lentzeallee 55, 14195 Berlin, Germany.

[3]21B Motijheel, Kolkata, West Bengal, India 700 074.

Correspondence should be addressed to Arunava Goswami agoswami@isical.ac.in

Malaria and other parasites, including virus often induce an increase in host lipids which the invaders use to their own advantage. We obtained encouraging results in our investigations on bird malaria with a new approach namely the use of nanosilica to mop up excess host lipids[1,2]. While this project is continuing we have investigated another, simpler system namely silkworms which suffer from a deadly baculovirus, BmNPV. This virus decimates the infected population within 24 hours or so and no known antibiotic antidote or genetically resistant strain of silkworm[3] exists. We report here a partial success, which is worth following up. Our rationale, we believe, has a broad and interdisciplinary appeal, for, this nanosilica treatment might be used together with other arsenals on all sorts of virus which take advantage of enhanced host lipids. It has not escaped our notice that Ebola and HIV also belong to this category. Nanoparticles are being preferentially harnessed, because they offer a greater surface area, circulate more easily and in lepidopteran system[4] they are removed within 24 hours from the body. Lawry[5] surmised, on cogent theoretical grounds that particles significantly smaller than micron order would be less harmful in the hemocoele. Furthermore, Hui-peng *et al.*[6] pointed out that lipase treatment, the only viable option for controlling BmNPV interferes in hormonal balance and cannot be applied to pre molting stage.

We tested a number of different nanosilica[7-9] with batches of 10 silkworms and the corresponding control. In future, the nanosilica in powder form will be dusted on mulberry leaves, the staple diet of *B. mori* silkworm. But at present nanosilica in ethanol (7 μg / μl; 5 μl per larva) has been injected in the laboratory. Initially, there was high mortality in controls injected with 5 μl ethanol but after practice the survival in 90 silkworms was 100%. Of the eight different nanosilicas (AL60101-AL60106, AL60110 and Advasan), all except Advasan permitted cocoon formation. However, with AL60106 cocoon formation was seen only in 35% of the worms. Prophylactic and pharmaceutical properties were most in AL60102, followed by AL60106. These nanosilicas are in the 50-60 nm size range with pore size range 3-10 nm.

In preliminary experiments we obtained ~80% survival after 24 hours following a single dose of AL60102, while all untreated virus infected larvae died within 24-30 hrs. We then tried with a larger number (70) of larvae (Table 1).

**Table 1 (a) Survival of the 5$^{th}$ instar silkworm after a single dose of AL60102; (b) survival at very late 5$^{th}$ instar larvae**

(a)

| Treatment | % survival after | |
|---|---|---|
| | 24 hours | 30 hours |
| Control (n=25) | 100 | 100 |
| Treated (n=70) | 85 | 75 |
| Virus (n=20) | 15* | 0 |

*morbid state

(b)

| Treatment | % survival after | |
|---|---|---|
| | 24 hours | 48 hours |
| Control (n=10) | 100 | 100 (all cocoons) |
| Treated (n=120) | 82* (2 cocoons) | ~65 (8 larvae, 21 cocoons) |
| Virus (n=20) | ~50 (1 cocoon) | ~ 35 (6 larvae, 1 cocoon) |

*Of these larvae, 44 were maintained for further observation, others being utilized for physiological and chemical experiments.

The increased viability about 65% over 35% in the larvae infected at the pre-molting stage is significant because lipase treatment[6] is inapplicable here. This relatively minor success might be enhanced with the help of feeding certain plant products[10] and mulberry leaves on which the nanosilica have been sprayed. Further injections are physiologically harmful according to our preliminary results.

In order to find out which lipid fractions, if any, are being removed by the nanosilica AL60102 and Al60106, TLC fractionation studies were attempted. Prefabricated HPTLC (Merck) on aluminium base were used. Haemolymph (30 µl) was directly spotted on HPTLC and run with the solvent system hexane: ether: acetic acid (90:10:1). The spots were visualized by putting the plates in iodine chamber and then immediately scanned. The relative differences in the intensities of TLC spots were measured by computer based pixel and area enumeration. The results consistently show that the most prominent spot of the normal haemolymph further significantly increases on infection with BmNPV. The compound that answers for the most prominent spot is presumably a liphophorin-lipid-lutein complex as reported by Tsuchida et al.[11]. Treatment of normal larva with AL60102 and AL60106, more so with AL60102, significantly increases this material (i.e., the intensity and area of the spot). Treatment of virus infected larvae with AL60102 and AL60106 significantly decreases this substance after some hours but not within 1 hour. After 6 hours the effect is marked and it is spectacular after ~24 hours. (The lipid content of uninfected haemolymph also increases on nanosilica treatment).

**Figure 1. AL60102 and AL60106 nearly totally abolish the protein-lipid-lutein complex. Sample spotted 24 hours after nanosilica treatment. N: Normal larvae without infection; N+102: AL60102 nanosilica injected in normal larvae; N+106: Another AL series nanosilica, AL60106 injected in normal larvae; V: BmNPV infected normal larvae; V+102: AL60102 nanosilica injected in virus infected larvae; V+106: AL60106 nanosilica injected in virus infected larvae.**

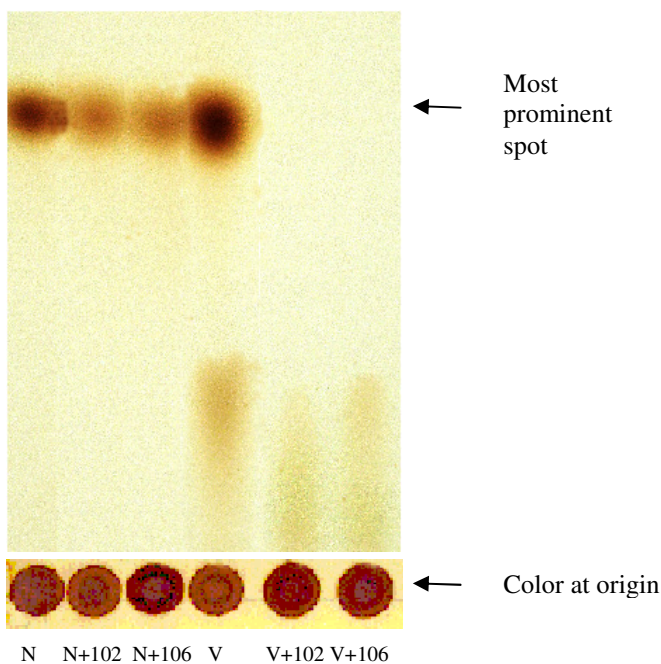

Figure 1 shows the spectacular results of AL60102 and AL 60106 exerted on the material in question after 24 hours. In fact, it almost vanishes (Table 2).

**Table 2. Pixel data on relative quantities of lipid spots 24 hours after AL60102 and AL60106 nanosilica treatment.**

| Figure1 spots | N | N+102 | N+106 | V | V+102 | V+106 |
|---|---|---|---|---|---|---|
| Top spot | 40.9 | 21.6 | 17.5 | 61.2 | 0.006 | 0.04 |
| Second spot | 75.04 | 75.31 | 79.26 | 77.64 | 79.34 | 72.19 |

Thus the enhanced survival and reduction of certain lipids (the top spot) by nanosilicas are apparently correlated.